\documentclass[sigconf]{acmart}

\usepackage{pifont}
\usepackage{enumitem}
\usepackage{subcaption}

\AtBeginDocument{%
  }

\copyrightyear{2026}
\acmYear{2026}
\setcopyright{cc}
\setcctype{by}
\acmConference[WWW '26]{Proceedings of the ACM Web Conference 2026}{April 13--17, 2026}{Dubai, United Arab Emirates}
\acmBooktitle{Proceedings of the ACM Web Conference 2026 (WWW '26), April 13--17, 2026, Dubai, United Arab Emirates}
\acmDOI{10.1145/3774904.3792851}
\acmISBN{979-8-4007-2307-0/2026/04}

\begin{document}

\title{DualGR: Generative Retrieval with Long and Short-Term \\ Interests Modeling}

\author{Zhongchao Yi}
\authornote{Work done during internship at Kuaishou Technology.}
\affiliation{%
  \institution{University of Science and Technology of China}
  \city{Hefei}
  \country{China}
}
\email{zhongchaoyi@mail.ustc.edu.cn}

\author{Kai Feng}
\affiliation{%
  \institution{Kuaishou Technology}
  \city{Beijing}
  \country{China}
}
\email{fengkai90@yeah.net}

\author{Xiaojian Ma}
\affiliation{%
  \institution{Kuaishou Technology}
  \city{Beijing}
  \country{China}
}
\email{maxiaojian@kuaishou.com}

\author{Yalong Wang}
\affiliation{%
  \institution{Kuaishou Technology}
  \city{Beijing}
  \country{China}
}
\email{wangyalong03@kuaishou.com}

\author{Yongqi Liu}
\affiliation{%
  \institution{Kuaishou Technology}
  \city{Beijing}
  \country{China}
}
\email{liuyongqi@kuaishou.com}

\author{Han Li}
\affiliation{%
  \institution{Kuaishou Technology}
  \city{Beijing}
  \country{China}
}
\email{lihan08@kuaishou.com}

\author{Zhengyang Zhou}
\authornote{Corresponding author.}
\affiliation{%
  \institution{University of Science and Technology of China}
  \city{Hefei}
  \country{China} \\
  \institution{Suzhou Institute for Advanced Research, University of Science and Technology of China}
  \city{Suzhou}
  \country{China}
}
\email{zzy0929@ustc.edu.cn}

\author{Yang Wang}
\affiliation{%
  \institution{University of Science and Technology of China}
  \city{Hefei}
  \country{China} \\
  \institution{Suzhou Institute for Advanced Research, University of Science and Technology of China}
  \city{Suzhou}
  \country{China}
}
\email{angyan@ustc.edu.cn}

\renewcommand{\shortauthors}{Zhongchao Yi et al.}

\begin{abstract}
In large-scale industrial recommendation systems, retrieval must produce high-quality candidates from massive corpora under strict latency.
Recently, Generative Retrieval (GR) has emerged as a viable alternative to Embedding-Based Retrieval (EBR), which quantizes items into a finite token space and decodes candidates autoregressively, providing a scalable path that explicitly models target-history interactions via cross-attention.
However, deploying GR in short-video feeds remains challenged by long-short interest interference, context-induced noise in hierarchical SID generation, and the lack of explicit learning from exposed-but-unclicked feedback.
To address these challenges, we propose \textbf{DualGR}, which combines (i) a Dual-Branch Long/Short-Term Router (DBR) with selective activation, (ii) Search-based \emph{SID} Decoding (S2D) that constrains fine-level decoding within the current coarse bucket for efficiency and noise control, and (iii) an Exposure-aware Next-Token Prediction Loss (ENTP-Loss) that treats unclicked exposures as coarse-level hard negatives to promote timely interest fade-out.
On the large-scale Kuaishou short-video recommendation system, DualGR has achieved outstanding performance. Online A/B testing shows +0.527\% video views and +0.432\% watch time lifts, validating DualGR as a practical and effective paradigm for industrial generative retrieval.
\end{abstract}




\maketitle

\begin{figure*}[!t]
  \vskip -0.10in
  \centering
  \includegraphics[width=0.98\linewidth]{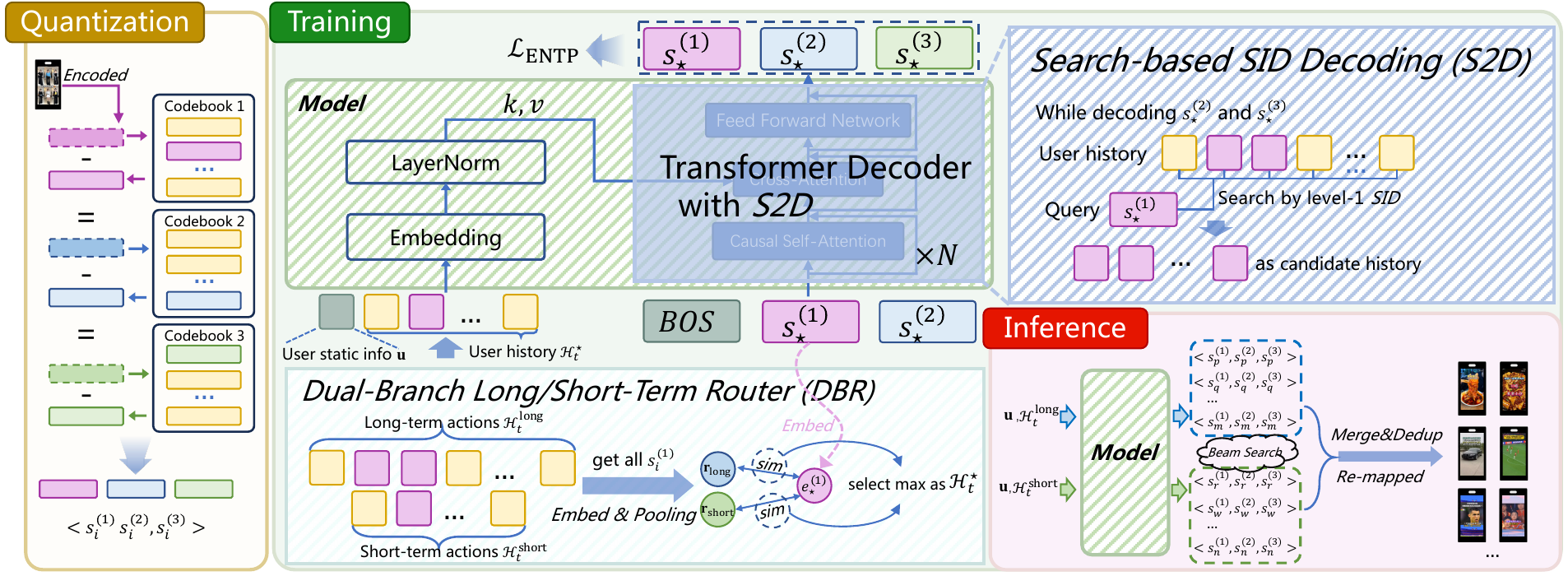}
  \vskip -0.10in
  \caption{Overview of the proposed DualGR.}
  \label{fig:framework}
  \vskip -0.15in
\end{figure*}

\section{Introduction}

In large-scale industrial recommender systems, retrieval is the first stage, which must select thousands of high-quality candidates from a multi-billion-item corpus for downstream ranking.
Our work is deployed in Kuaishou's double-column Explore feed, akin to Instagram's two-column waterfall layout, where multiple cards appear on the same screen. Compared with the single-column full-screen feed, this layout places stricter demands on candidate diversity, which must produce diverse yet relevant candidates in real time to cover multiple topics and intents, also to rapidly track emerging interests, therefore it magnifies modeling and efficiency challenges at the retrieval stage.
In industry, the dominant approach is Embedding-Based Retrieval (EBR), whose DNN twin-tower architecture retrieves user-item matches via approximate nearest neighbors (ANN) ~\cite{meng2025user,liu2024kuaiformer}.
While highly scalable, EBR is inherently limited in multi-interest modeling and target-aware interaction, and is vulnerable to exposure bias~\cite{cen2020controllable,yi2019sampling}.

Recently, inspired by the generative progress of LLMs and Transformers~\cite{zhao2023survey}, industry has begun adopting Generative Retrieval (GR), whose items are discretized into multi-level \emph{semantic IDs (SIDs)} and candidates are decoded autoregressively along the hierarchy~\cite{rajput2023recommender,yang2024unifying,zhou2025onerec}.
Operating within the same finite and discrete token space during both training and inference can effectively alleviate exposure bias; moreover, cross-attention in decoding enables explicit target-history interaction. Combined with beam search, the model can naturally produce diverse candidate sets under comparable latency budgets, aligning with retrieval's coverage and exploration goals.
Despite this promise, deploying the generative paradigm robustly in multi-interest, short-video retrieval still faces three key \textbf{\textit{challenges}}:
\ding{182} \textbf{Long-short interest interference.} The double-column feed is highly sensitive to both persistent preferences and hotspot tracking; existing "multi-interest" effects often emerge passively via multi-beam decoding and lack explicit control across time scales. When stable preferences and transient hotspots coexist, attention and gradients can dilute each other, destabilizing routing and decoding and upsetting the relevance-diversity balance.
\ding{183} \textbf{Context noise and long-history constraints.} During fine-grained (level-2/3) \emph{SIDs} generation, the model struggles to separate target-relevant fragments from irrelevant noise, akin to context engineering issues observed in LLMs~\cite{liu2023lost}. In the same time,  under the strict latency budgets of retrieval, without category-wise coarse screening before interaction, the use of long histories will become prohibitive~\cite{pi2020search, meng2025user}.
\ding{184} \textbf{Missing negative feedback.} Industrial logs contain many unclicked exposed items, yet in the next-token prediction (NTP)~\cite{rajput2023recommender, zhou2025onerec}, it's difficult to encode them as explicit negative signals, impeding the timely fade-out of non-interest directions and thereby degrading decoding quality and coverage efficiency.

To address these issues, we propose \textbf{DualGR}, a generative retrieval framework with two explicit interest branches and search-based decoding, while enabling the timely fade-out of non-interest.
First, we utilizes \emph{Dual-Branch Long/Short-Term Router (DBR)} to decompose a user's history into long-term and short-term summaries, compute target-aware similarities, and route to exactly one branch to suppress cross-branch dilution; at inference, run one forward pass per branch and merge the candidates to \textbf{cover both stable preferences and transient intents}.
Second, we introduce \emph{Search-based SID Decoding (S2D)},
when predicting fine-grained (level-2/3) \emph{SIDs}, candidate actions are constrained to the current coarse (level-1) bucket, which shrinks the search space and reinforces intra-bucket consistency, thereby \textbf{enabling longer usable histories while reducing noise interference and improving computational efficiency}.
Third, to overcome the lack of negatives, we propose
\emph{Exposure-aware Next-Token Prediction Loss (ENTP-Loss)}, at level-1, we treat unclicked exposures as hard negatives, \textbf{explicitly penalizing the corresponding coarse categories to accelerate the fade-out of non-interest directions}.
Considering latency and compute budgets, we adopt an encoder-free lightweight frontend (embeddings + LayerNorm) and perform step-wise cross attention over the action list,
achieving SIM-like target awareness without redundant re-encoding~\cite{pi2020search}.
The \textbf{\textit{contributions}} are as follows:
\begin{itemize}[leftmargin=*]
\item We propose \textbf{DualGR}, a generative retriever that explicitly models users' long- and short-term interests while enabling the timely fade-out of non-interest, providing a practical and effective paradigm for industrial generative retrieval.
\item We design \emph{Dual-Branch Long/Short-Term Router (DBR)} to capture users' long- and short-term preferences, \emph{Search-based SID Decoding (S2D)} to suppress context-induced noise, and improve computational efficiency; and \emph{Exposure-aware Next-Token Prediction (ENTP-Loss)} to bring in unclicked exposures for timely fade-out of non-interest. Taken together, these components make retrieval more interpretable and controllable.
\item Extensive experiments on Kuaishou's double-column Explore feed short-video recommendation with hundreds of millions of users, demonstrates the effectiveness of DualGR.
\end{itemize}

\section{Methodology}

\subsection{Preliminary}

In short-video recommendation, we model user interests from past interactions and predict which videos the user is likely to engage with next. Let $\mathcal{V}$ be the universe of videos, for a user $u$ at request time $t$, the context consists of a static user feature $\mathbf{u}$ (\textit{e.g.}, age, gender) and a chronologically ordered action history $\mathcal{H}_t = (\mathbf{a}_1,\ldots,\mathbf{a}_{T_t})$, where each action $\mathbf{a}_j$ includes the video ID $v_j \in \mathcal{V}$, author ID, interaction tags (\textit{e.g.}, like, follow), and watch time, etc. The goal of retrieval stage is to return a candidate set 
\begin{equation}
\mathcal{R}_{\theta}(\mathbf{u},\mathcal{H}_t)
~=~
\text{argmax}
~U\!\big(\mathcal{R_{\theta}};\mathbf{u},\mathcal{H}_t  \big).
\label{eq:retri}
\end{equation}
where $\mathcal{R}_{\theta}(\mathbf{u},\mathcal{H}_t) = \{v_1,v_2,...,v_k\} \subseteq \mathcal{V}$, \(U(\cdot)\) denotes the overall online-utility objective (\textit{e.g.}, watch time, video views, diversity).

For generative retrieval, each video $v \in \mathcal{V}$ is discretized into hierarchical \emph{semantic IDs (SIDs)}: $
\big(s^{(1)},\, s^{(2)},\, \ldots,\, s^{(L)}\big)$, where $s^{(\ell)} \in \mathcal{C}^{(\ell)}$, $\mathcal{C}^{(\ell)}$ is the level-$\ell$ codebook  ~\cite{rajput2023recommender,yang2024unifying,zhou2025onerec}.
A generative retriever defines a conditional distribution over hierarchical \emph{SIDs} via:
\begin{equation}
p_\theta\!\big(s^{(1:L)} \mid x_t\big)
~=~
\prod_{\ell=1}^{L}
p_\theta\!\big(s^{(\ell)} ~\big|~ s^{(1:\ell-1)},\, x_t\big),
\end{equation}
where $x_t$ is encoded from $(\mathbf{u},\mathcal{H}_t)$. The training objective maximizes the likelihood of the ground-truth multi-level \emph{SID}.

At serving time, the Top-$K$ \emph{SIDs} are decoded under 
beam search (beam size $B$) or sampling~\cite{zhou2025onerec}: $\widehat{\mathcal{S}}_{\theta}(x_t)=\operatorname{TopK}_B\Big\{p_\theta\!\big(s^{(1:L)}\!\mid x_t\big)\Big\}.$
Finally, a prebuilt mapping/index $\mathrm{Map}: \mathcal{S} \rightarrow \mathcal{V}$ projects \emph{SIDs} back to concrete video ids forming the retrieval result $ \mathcal{R}_{\theta}(x_t) = \mathrm{Map}\!\big(\widehat{\mathcal{S}}_{\theta}(x_t)\big) \subseteq \mathcal{V}$, which is also required to satisfy Eq.~\ref{eq:retri}.

\subsection{Framework Overview}
We first pre-quantize videos into \(L{=}3\) hierarchical \emph{SIDs} via residual k-means (RQ-KMeans)~\cite{zhou2025onerec}.
Built on an encoder-free, decoder-centric backbone, \textbf{DualGR} integrates three components, \textit{i.e.}, \textbf{DBR} (long/short routing), \textbf{S2D} (search-based fine decoding), and \textbf{ENTP-Loss} (exposure-aware training) as illustrated in Fig.~\ref{fig:framework}.

\subsection{Dual-Branch Long/Short-Term Router}
\label{sec:dbr}
To explicitly capture users' long- and short-term dependencies, we propose the Dual-Branch Long/Short-Term Router (DBR), which defines two windows 
\(
\mathcal{H}^{\text{long}}_t
\) (length $L_{\text{long}}$) and
\(
\mathcal{H}^{\text{short}}_t
\) (length $L_{\text{short}}<L_{\text{long}}$),
obtained from $\mathcal{H}_t$ by taking the most recent actions with padding if needed.
Then DBR maps video's level-1 \emph{SID} in each action to an embedding using the  vocabulary embedding $E_{vocab}^{(1)}\!\in\!\mathbb{R}^{|\mathcal{C}^{(1)}|\times d}$ and form two target-aware summaries via pooling:
\begin{align}
\mathbf{r}_{\text{long}} \!&=\! \mathrm{Pool}\big(\{E_{vocab}^{(1)}[s^{(1)}(\mathbf{a})]:\mathbf{a}\!\in\!\mathcal{H}^{\text{long}}_t\}\big)
\\
\mathbf{r}_{\text{short}} \!&=\! \mathrm{Pool}\big(\{E_{vocab}^{(1)}[s^{(1)}(\mathbf{a})]:\mathbf{a}\!\in\!\mathcal{H}^{\text{short}}_t\}\big)
\label{eq:dbr-pool}
\end{align}
For a training example with ground-truth level-1 token $s_\star^{(1)}$ (embedding $\mathbf{e}_\star^{(1)}{=}E_{vacob}^{(1)}[s_\star^{(1)}]$), DBR computes target-aware similarities:
\begin{equation}
\gamma_{\text{long}}=\cos(\mathbf{e}_\star^{(1)},\mathbf{r}_{\text{long}}),\,
\gamma_{\text{short}}=\cos(\mathbf{e}_\star^{(1)},\mathbf{r}_{\text{short}}),
\label{eq:dbr-sim}
\end{equation}
applying a hard gate to select the window with the larger similarity as the coarse-step user history $\mathcal{H}^\star_t$.
Subsequently, DBR constructs the input by concatenating it with user static features, then applying embeddings and LayerNorm:
\begin{equation}
x^\star_t=\mathrm{LN}\Big(\big[\,E_u(\mathbf{u})\;;\;E_a\big(\mathcal{H}^\star_t\big)\,\big]\Big)
\label{eq:encode}
\end{equation}
where $E_u$ and $E_a$ are embedding transforms for user statics and actions.
The decoder conditions on $x^\star_t$ to predict the level-1 token $s^{(1)}$, which avoids long-short term dilution and yields a cleaner supervision signal for the coarse category.

At inference time, DBR instantiates both branches
\begin{equation}
x^\text{long}_t=\mathrm{LN}\big([E_u(\mathbf{u});E_a(\mathcal{H}^{\text{long}}_t)]\big),\,
x^\text{short}_t=\mathrm{LN}\big([E_u(\mathbf{u});E_a(\mathcal{H}^{\text{short}}_t)]\big),
\end{equation}
and runs one beam-search decoding pass per branch to obtain
the coarse level-1 tokens. Following the fine-grained decoding in Sec.~\ref{sec:s2d}, each branch continues beam search to decode $\hat{s}^{(2)},\hat{s}^{(3)}$ and form a candidate \emph{SID} set.
Then the two sets are merged as $\widehat{\mathcal{S}}_{\theta}(x_t)$,
and finally mapped back to concrete videos to produce the retrieval set, thereby jointly covering both long- and short-term interests.

\subsection{Search-based \emph{SID} Decoding}
\label{sec:s2d}
For fine-grained decoding ($\ell\!>\!1$), Search-based \emph{SID} Decoding (S2D) ignores the long/short windowing and searches over the original action list $\mathcal{H}_t$, so as to harvest as many in-bucket interactions as possible once the coarse category is determined.
\subsubsection{Training}
Let $s_\star^{(1)}$ denote the coarse (level-1) \emph{SID} of the training target.
Given $\mathcal{H}_t=(\mathbf{a}_1,\ldots,\mathbf{a}_{T_t})$, S2D constructs a same-bucket, recent subset via
\begin{equation}
\widehat{\mathcal{H}}_t(s_\star^{(1)})
~=~
\mathrm{Search}(\{\mathbf{a}\in\mathcal{H}_t \;:\; s^{(1)}(\mathbf{a}) = s_\star^{(1)}\}).
\label{eq:s2d-hist}
\end{equation}
then builds the fine-level input by concatenating user static features with the selected actions and encoding them as in Eq.~\ref{eq:encode} to obtain $x_t^{(>1)}$. Under teacher forcing~\cite{zhao2023survey}, the conditional likelihood for the remaining tokens is
\begin{equation}
p_\theta\!(s_\star^{(\ell)} \,\big|\, s_\star^{(1:\ell-1)}, \;x_t^{(>1)}),\;\ell = 2,\ldots,L
\end{equation}
, which is optimized by minimizing the ENTP objective in Sec.~\ref{sec:entp-loss}.

\subsubsection{Inference}
At serving, each branch first predicts $\hat{s}^{(1)}$ as described in Sec.~\ref{sec:dbr}. Subsequently, S2D runs the global search on the full action list $\mathcal{H}_t$ to obtain $\widehat{\mathcal{H}}_t(\hat{s}^{(1)})$ via Eq.~\eqref{eq:s2d-hist}, forming $x^{(\ge 1)}$, and then decodes $s^{(2)},s^{(3)}$ autoregressively.

S2D realizes the decoding with searching, controlled noise and latency. Conditioning on the chosen level-1 bucket $s_\star^{(1)}$, the size of the in-bucket history scales approximately as $ \big|\mathcal{H}_t\!\big(s_\star^{(1)}\big)\big|\approx \frac{|\mathcal{H}_t|}{|\mathcal{V}^{(1)}|}$. In practice, the codebook per level is typically large (\textit{e.g.}, $\ge 256$ and up to $8192$)~\cite{rajput2023recommender,zhou2025onerec}, hence $\big|\mathcal{H}_t(s_\star^{(1)})\big|\!\ll\!|\mathcal{H}_t|$. This shrinkage enables longer usable histories under a fixed latency budget.

\subsection{Exposure-aware NTP Loss}
\label{sec:entp-loss}
To encourage timely fade-out of non-interest directions, we propose an Exposure-aware NTP Loss (\textbf{ENTP-Loss}).
Specifically, we organize training tuples as $(x_i, s_i^{(1:L)}, c_i)$ where $c_i\!\in\!\{0,1\}$ indicates a \emph{positive} (valid interactive) $s_i^{(1:L)}$ if $c_i{=}1$, and $c_i{=}0$ for the \textit{negative} (unclicked exposure) $s_i^{(1:L)}$.
Let \(p_i^{(\ell)}\) be the predicted probability (clipped to \([\varepsilon,1-\varepsilon]\)) of the ground-truth \emph{SID} at level \(\ell\), 
the ENTP-Loss can be defined as 
\begin{equation}
\mathcal{L}_{\text{ENTP}}{=}\frac{1}{N}\!\sum_{i\in\mathcal{B}}\!
\left[ c_i \sum_{\ell=1}^{L}\!\big(-\!\log p_i^{(\ell)}\big)\!+\!
(1\!-\!c_i)\big(\!-\!\alpha\log\!\big(1\!-\!p_i^{(1)}\big)\big)\right]
\label{eq:entp-batch}
\end{equation}
where \(\mathcal{B}\) is a mini-batch of size \(N{=}\lvert\mathcal{B}\rvert\) and \(\alpha\!>\!0\) controls the strength of the negative signal. This objective keeps the standard NTP likelihood on positives at all levels while adding a coarse-level (level-1) penalty for negatives, 
promoting timely non-interest fade-out without introducing fine-level label noise.

\section{Experiments}

\begin{figure}[h]
  \vskip -0.1in
  \centering
  \begin{subfigure}{0.48\linewidth}
    \centering
    \includegraphics[width=\linewidth]{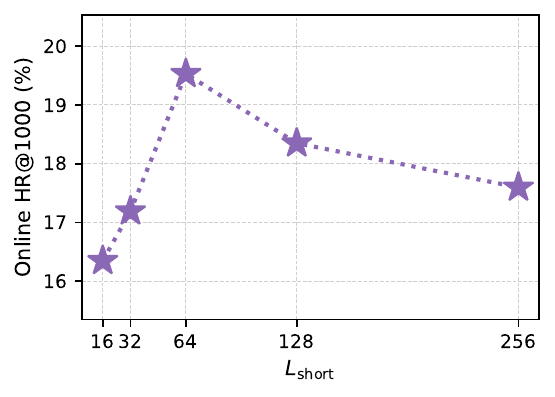}
    \vskip -0.05in
    \caption{Short-term window length.}
    \label{fig:shortlen}
  \end{subfigure}\hfill
  \begin{subfigure}{0.48\linewidth}
    \centering
    \includegraphics[width=\linewidth]{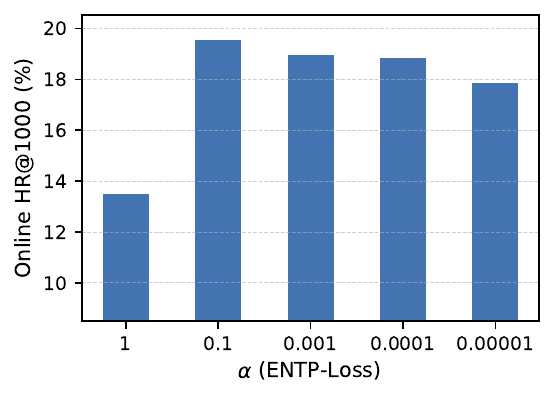}
    \vskip -0.05in
    \caption{$\alpha$ in ENTP-Loss.}
    \label{fig:alpha}
  \end{subfigure}
  \vskip -0.10in
  \caption{Sensitivity analysis of DualGR.}
  \label{fig:para}
  \vskip -0.2in
\end{figure}

\textit{\textbf{Experimental Settings.}}
We conduct experiments on Kuaishou's double-column Explore feed short-video stream, one of the platform's largest recommendation surfaces, serving over \textbf{100M users, 1B video views, and 10B exposures} per day. We compare DualGR against strong retrieval baselines deployed in this scenario, including \textbf{ComiRec}~\cite{cen2020controllable}, \textbf{PDN}~\cite{li2021path}, and \textbf{KuaiFormer}~\cite{liu2024kuaiformer}; additionally, for a fair comparison with generative retrieval, we implement \textbf{TIGER}~\cite{rajput2023recommender}. For evaluation, we use \textbf{Online Hit Rate (HR)}, a metric widely adopted in prior work~\cite{liu2024kuaiformer}.
\textbf{DualGR} adopts an encoder-free decoder of 4 Transformer blocks with model dimension 512, 8 attention heads, and feed-forward dimension 1024.
Videos are pre-quantized into $L{=}3$ hierarchical codebooks with per-level size $|\mathcal{C}^{(\ell)}|{=}8192$.
For inputs, we use the user's most recent positive interactions where $|\mathcal{H}_t|{=}1000$, the DBR employs window lengths $L_{\text{long}}{=}1000$ and $L_{\text{short}}{=}64$.
Training uses the ENTP-Loss with $\alpha{=}0.1$.
\\
\textit{\textbf{Overall Performance.}}
Table~\ref{tab:overall} summarizes the results on the short-video recommendation. DualGR consistently outperforms strong baselines and the implemented generative retriever, demonstrating clear advantages. These gains indicate that explicitly modeling long- and short-term interests with search-based activation, coupled with fade-out of non-interest, yields a more effective and robust generative retrieval paradigm for industrial short-video feeds.
\\
\textit{\textbf{Sensitivity Analysis.}}
We study the impact of the short-term window length $L_{\text{short}}$ (Fig.~\ref{fig:shortlen}). Small windows underfit recency, missing short-term interests, while large windows blur the boundary with the long-term branch, so a moderate $L_{\text{short}}$ can best captures recency and remains complementary.
For $\alpha$ in ENTP-Loss (Fig.~\ref{fig:alpha}), overly large values over-penalize unclicked exposures, thus destabilizing training and suppressing true interests, whereas overly small values yield slow fade-out. A moderate setting strikes the best balance.

\begin{table}[t]
\vskip -0.1in
\centering
\setlength{\tabcolsep}{8pt}
\caption{Overall online HR@K performance.}
\label{tab:overall}
\vskip -0.15in
\begin{tabular}{lccc}
\toprule
\textbf{Method} & \textbf{HR@100} & \textbf{HR@500} & \textbf{HR@1000} \\
\midrule
ComiRec         &        2.539\%         &     6.134\%            &      8.249\%            \\
PDN             &        2.343\%         &     5.483\%            &      7.249\%            \\
KuaiFormer      &        4.495\%         &     7.251\%            &      12.356\%            \\
TIGER           &        4.936\%         &     8.184\%            &      14.442\%            \\
\midrule
\textbf{DualGR} &    \textbf{6.827\%}    &   \textbf{10.319\%}    &    \textbf{19.529\%} \\
\bottomrule
\end{tabular}
\vskip -0.15in
\end{table}

\subsection{Online A/B Test}
We conducted a week-long online A/B experiment on Kuaishou's double-column Explore feed with 6\% traffic,
integrating DualGR as an additional retrieval channel. Results show consistent lifts of \textbf{+0.527\%} video views and \textbf{+0.432\%} watch time, while keeping end-to-end latency within service-level budgets (no measurable response time increase), validating DualGR's practical effectiveness.

\subsection{Ablation Study}
We assess the contribution of each module via three variants (Table~\ref{tab:ablation}), \textit{i.e.}, \textbf{w/o DBR} replaces the DBR with long branch only, \textbf{w/o S2D} removes S2D and performs unconstrained fine-level decoding, and \textbf{w/o ENTP-Loss} trains with vanilla NTP without exposure-aware negatives.
DualGR consistently outperforms all variants verifying that all components are necessary and complementary.

\begin{table}[t]
\vskip -0.1in
\centering
\setlength{\tabcolsep}{8pt}
\caption{Ablation study of DualGR.}
\label{tab:ablation}
\vskip -0.15in
\begin{tabular}{lccc}
\toprule
\textbf{Method} & \textbf{HR@100} & \textbf{HR@500} & \textbf{HR@1000} \\
\midrule
w/o DBR         &      5.134\%           &      8.892\%           &   15.379\%     \\
w/o S2D         &      6.257\%           &      9.182\%           &   16.287\%     \\
w/o ENTP-Loss   &      6.672\%           &      9.837\%           &   17.576\%     \\
\midrule
\textbf{DualGR} &    \textbf{6.827\%}    &   \textbf{10.319\%}    &    \textbf{19.529\%} \\
\bottomrule
\end{tabular}
\vskip -0.15in
\end{table}

\section{Conclusion}
We present \textbf{DualGR}, a generative retriever that explicitly models long- and short-term interests via a Dual-Branch Router (DBR), controls context noise through Search-based \emph{SID} Decoding (S2D), and accelerates non-interest fade-out with an Exposure-aware NTP Loss (ENTP-Loss). Built on an encoder-free, step-wise target-aware decoder, DualGR delivers consistent HR@K gains and strong online lifts in our production setting, demonstrating its effectiveness as a practical paradigm for industrial generative retrieval.

\section*{Acknowledgements}
This paper is partially supported by the National Natural Science Foundation of China (No.62502488, No.12227901), Natural Science Foundation of Jiangsu Province (BK20240460), the grant from State Key Laboratory of Resources and Environmental Information System.

\bibliographystyle{ACM-Reference-Format}
\bibliography{DualGR}

@inproceedings{cen2020controllable,
  title={Controllable multi-interest framework for recommendation},
  author={Cen, Yukuo and Zhang, Jianwei and Zou, Xu and Zhou, Chang and Yang, Hongxia and Tang, Jie},
  booktitle={Proceedings of the 26th ACM SIGKDD international conference on knowledge discovery \& data mining},
  pages={2942--2951},
  year={2020}
}

@article{zhou2025onerec,
  title={OneRec Technical Report},
  author={Zhou, Guorui and Deng, Jiaxin and Zhang, Jinghao and Cai, Kuo and Ren, Lejian and Luo, Qiang and Wang, Qianqian and Hu, Qigen and Huang, Rui and Wang, Shiyao and others},
  journal={arXiv preprint arXiv:2506.13695},
  year={2025}
}

@inproceedings{yi2019sampling,
  title={Sampling-bias-corrected neural modeling for large corpus item recommendations},
  author={Yi, Xinyang and Yang, Ji and Hong, Lichan and Cheng, Derek Zhiyuan and Heldt, Lukasz and Kumthekar, Aditee and Zhao, Zhe and Wei, Li and Chi, Ed},
  booktitle={Proceedings of the 13th ACM conference on recommender systems},
  pages={269--277},
  year={2019}
}

@inproceedings{pi2020search,
  title={Search-based user interest modeling with lifelong sequential behavior data for click-through rate prediction},
  author={Pi, Qi and Zhou, Guorui and Zhang, Yujing and Wang, Zhe and Ren, Lejian and Fan, Ying and Zhu, Xiaoqiang and Gai, Kun},
  booktitle={Proceedings of the 29th ACM International Conference on Information \& Knowledge Management},
  pages={2685--2692},
  year={2020}
}

@article{rajput2023recommender,
  title={Recommender systems with generative retrieval},
  author={Rajput, Shashank and Mehta, Nikhil and Singh, Anima and Hulikal Keshavan, Raghunandan and Vu, Trung and Heldt, Lukasz and Hong, Lichan and Tay, Yi and Tran, Vinh and Samost, Jonah and others},
  journal={Advances in Neural Information Processing Systems},
  volume={36},
  pages={10299--10315},
  year={2023}
}

@article{yang2024unifying,
  title={Unifying generative and dense retrieval for sequential recommendation},
  author={Yang, Liu and Paischer, Fabian and Hassani, Kaveh and Li, Jiacheng and Shao, Shuai and Li, Zhang Gabriel and He, Yun and Feng, Xue and Noorshams, Nima and Park, Sem and others},
  journal={arXiv preprint arXiv:2411.18814},
  year={2024}
}

@article{zhao2023survey,
  title={A survey of large language models},
  author={Zhao, Wayne Xin and Zhou, Kun and Li, Junyi and Tang, Tianyi and Wang, Xiaolei and Hou, Yupeng and Min, Yingqian and Zhang, Beichen and Zhang, Junjie and Dong, Zican and others},
  journal={arXiv preprint arXiv:2303.18223},
  volume={1},
  number={2},
  year={2023}
}

@article{liu2023lost,
  title={Lost in the middle: How language models use long contexts},
  author={Liu, Nelson F and Lin, Kevin and Hewitt, John and Paranjape, Ashwin and Bevilacqua, Michele and Petroni, Fabio and Liang, Percy},
  journal={arXiv preprint arXiv:2307.03172},
  year={2023}
}

@inproceedings{meng2025user,
  title={User Long-Term Multi-Interest Retrieval Model for Recommendation},
  author={Meng, Yue and Guo, Cheng and Hu, Xiaohui and Deng, Honghu and Cao, Yi and Liu, Tong and Zheng, Bo},
  booktitle={Proceedings of the Nineteenth ACM Conference on Recommender Systems},
  pages={1112--1116},
  year={2025}
}

@article{liu2024kuaiformer,
  title={KuaiFormer: Transformer-Based Retrieval at Kuaishou},
  author={Liu, Chi and Cao, Jiangxia and Huang, Rui and Zheng, Kai and Luo, Qiang and Gai, Kun and Zhou, Guorui},
  journal={arXiv preprint arXiv:2411.10057},
  year={2024}
}

@article{li2021path,
  title={Path-based Deep Network for Candidate Item Matching in Recommenders},
  author={Li, Houyi and Chen, Zhihong and Li, Chenliang and Xiao, Rong and Deng, Hongbo and Zhang, Peng and Liu, Yongchao and Tang, Haihong},
  journal={arXiv preprint arXiv:2105.08246},
  year={2021}
}

\end{document}